# Matter and spin superposition in vacuum experiment (MASSIVE)

Sougato Bose[1] and Gavin W. Morley[2,*]

[1] Department of Physics and Astronomy, University College London, Gower Street, London WC1E 6BT, United Kingdom

[2] Department of Physics, University of Warwick, Gibbet Hill Road, Coventry CV4 7AL, United Kingdom

* gavin.morley@warwick.ac.uk

A free-falling nanodiamond containing a nitrogen vacancy centre in a spin superposition should experience a superposition of forces in an inhomogeneous magnetic field. We propose a practical design that brings the internal temperature of the diamond to under 10 K. This extends the expected spin coherence time from 2 ms to 500 ms, so the spatial superposition distance could be increased from 0.05 nm to over 1 µm, for a 1 µm diameter diamond and a magnetic inhomogeneity of only $10^4$ T/m. The low temperature allows single-shot spin readout, reducing the number of nanodiamonds that need to be dropped by a factor of 10,000. We also propose solutions to a generic obstacle that would prevent such macroscopic superpositions.

Cats have not been observed in a quantum superposition [1] but this is routine for atoms [2]. Between these two extremes there may be a quantum-to-classical transition which could solve the quantum measurement problem: why is the deterministic evolution of the Schrödinger equation sometimes interrupted by a measurement? One concrete proposal for a quantum-to-classical transition is continuous spontaneous localization (CSL) [3, 4], and this leads to a measure of how macroscopic a superposition is, based on how well it would exclude CSL [5]. The most macroscopic superposition created so far by this measure is matter-wave interference of molecules made of over 800 atoms going through gratings with period 266 nm [6, 7].

One set of proposals for a more macroscopic superposition is based on levitated nanodiamonds or microdiamonds containing a single nitrogen vacancy centre ($NV^-$) [8-13]. The core feature of these schemes is that the NV- is put into an electron spin superposition so that an inhomogeneous magnetic field creates a superposition of forces and hence a spatial superposition. The two components of this matter wave are then interfered with each other to produce fringes. One of our proposals introduced the idea of a freely falling nanodiamond in order to reach greater spatial superposition distance by avoiding the trapping force which tends to hold the two spatial superposition components together [11]. Like our earlier proposal, it is not necessary to cool the centre of mass motion of the nanodiamond to the quantum ground state because the spin is in a pure state [9]. Experiments making progress in this area include nanodiamonds levitated with optical tweezers [14-19], ion-traps [20-24] and a magnetogravitational trap [25]. The rotational motion of levitated nanodiamonds has also been shown to be relevant in proposals [26] and experiments [27].

The optimum diameter for the diamond in these proposals is around 1 µm which corresponds to around 100 billion atoms with a mass of over $10^{-15}$ kg. A 1 µm sphere has most of its volume more than 100 nm from the surface, where the NV- spin coherence times should be similar to bulk values. Microdiamonds are not spherical [19] which will tend to bring the NV- closer to the surface: it will be necessary to select diamonds containing a single NV- towards the centre so as to have bulk-like spin and optical coherence.

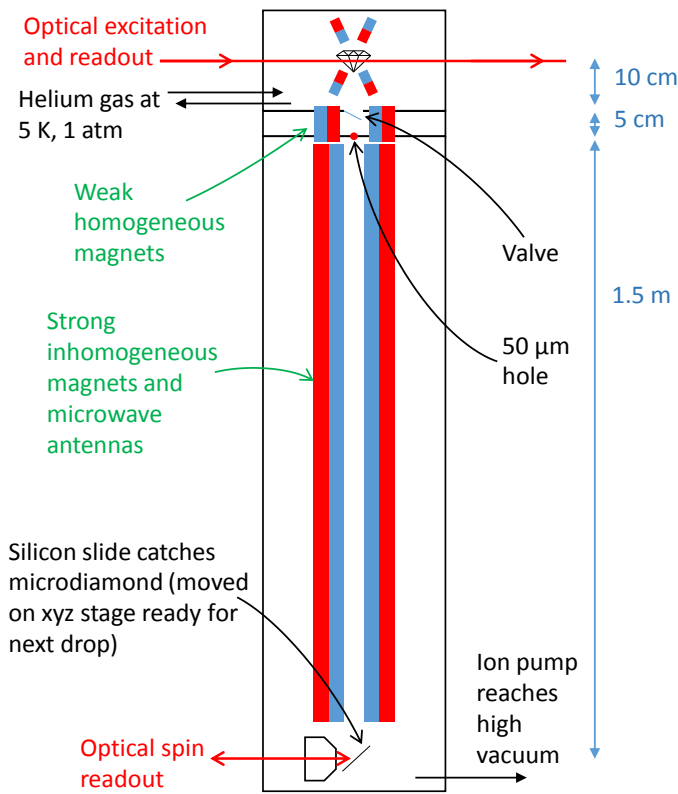

Figure 1. Schematic of the proposed experiment for testing macroscopic superpositions. A microdiamond is levitated at the top in a magnetogravitational trap, before being cooled with 5 K helium gas and then dropped through the ultra-high vacuum (UHV) chamber within an inhomogeneous magnetic field. Optical spin readout is performed after the drop. The magnets and the silicon slide are held at 4.2 K. Magnets are shown schematically with their north (south) poles in blue (red). The diamond and 50 μm hole are not to scale as they would be too small to see.

Our proposal is shown schematically in Fig. 1: a key feature is that after trapping a microdiamond, cold helium gas should be introduced to lower the internal temperature. After this is pumped away, the diamond will remain cold if the trap is a magnetogravitational trap [25] or an ion trap [20-24]. We have shown that optical traps heat impure (commercially-available) nanodiamonds severely [18], and even custom-made high-purity nanodiamonds will absorb enough of the high-power trapping light to warm above 10 K [19]. To have a single NV- centre inside of a 1 μm diamond, we could use 'low absorption' CVD diamond as a starting material for milling of microdiamonds [19]. This has around 20 ppb nitrogen impurities, so there would be 1800 nitrogen impurities in each microdiamond and hence around six grown-in NV- per diamond [28] with no need to add vacancies. Grown-in NV- are known to have spin coherence times, $T_2$, that are as long or longer than NV- that are made deliberately. There are four possible NV- orientations, so with around six NV- it is very likely that there will be one orientation for which there is just one NV-. The other NV- centres can then be ignored because they will be spin polarized to zero and then their spin will be unaffected by the microwave pulses that follow which will be on resonant only with the chosen orientation.

Isotopically-purified $^{12}$C diamond with CPMG dynamic decoupling [29] or dynamic decoupling based on measurements of the spin environment [30] could be used to reach electron spin coherence times of over 0.5 s. The CPMG pulse sequence is $(\pi/2)_x$ [ – $(\pi)_y$ ]$_n$ where x and y are the axes in the Bloch sphere about which the spin is rotated and n is the number of times that the $(\pi)_y$ pulse is repeated after a delay period. This sequence would not prevent the electron spin from generating the spatial superposition as the $(\pi)_y$ pulses keep the spin pointing along the y axis. Most dynamic decoupling sequences do not have this feature: some of them are designed to deal with arbitrary initial spin states which is not required here.

The full experimental protocol would run as follows. (1) Trap a microdiamond and reduce the pressure to some convenient level around 5 mbar so that the diamond mass can be measured by fitting the power-spectral density [31, 32]. Electrically neutralise the diamond with a radioactive source and UV light [25, 33, 34]. (2) Collect an optical florescence spectrum, to confirm the presence of NV-. (3) Perform a Hanbury Brown-Twiss experiment to check how many NV- are in the diamond: 532 nm light excites the NV- and the timing of the collected florescence can be analysed to measure the second-order autocorrelation function, $g^{(2)}(t)$. For $g^{(2)}(t=0) < 0.5$ we can infer that a single NV- is present. (4) Measure optically-detected magnetic resonance (ODMR) spectrum to check the alignment of the NV-. (5) Admit one atmosphere of helium gas at 4.2 K to the vacuum chamber to cool the microdiamond, and then reduce the pressure to around 100 mbar leaving the internal temperature of the diamond around 5 K. (6) Optically polarize the NV- spin, reduce the pressure to $10^{-6}$ mbar or lower and open a valve so as to drop the diamond through a 50 µm hole into an ultra-high vacuum (UHV) chamber. The pressure in the UHV chamber would be maintained by an ion pump or a diffusion pump, with a large pressure difference across the 50 µm hole. A homogeneous magnetic field of around 50 mT would be used to keep the NV- oriented with the inhomogeneous magnetic field below. (7) Apply a microwave π/2 pulse to change the spin state from $|0\rangle$ to $\frac{1}{\sqrt{2}}(|0\rangle + |1\rangle)$. It would be possible to reach spatial superposition distances twice as large by using the double quantum state $\frac{1}{\sqrt{2}}(|-1\rangle + |1\rangle)$ instead [35], but this would increase the number of pulses required for the dynamic decoupling, which would make the decoupling less efficient, reducing the spin coherence time. (8) An inhomogeneous magnetic field is present throughout the 1.5 m drop which lasts 0.4 s. (9) A time $t_1$ after the creation of the superposition, flip the spin state with a microwave $\pi_x$ pulse so that the matter wave components are brought back together, and then a time $t_2$ after the creation of the superposition, apply another $\pi_x$ pulse to slow down the diamond before the interference at time $t_3$. This sequence was proposed in reference [11]. (10) Catch the diamond on a silicon slide which is held at 5 K and read out the spin state optically. (11) Move the silicon slide slightly and repeat from step 1 with slightly different values of the drop time to search for fringes. (12) Measure the (dynamic decoupling) spin coherence time of each of the NV- in different microdiamonds in a row on the silicon slide.

**More detailed considerations about the above steps**

The diamonds trapped will have a range of masses which provides a valuable data set as a function of mass. It is possible that CSL or some other mechanism prevents spatial superpositions for the more massive particles, while they are allowed for the less massive particles.

The initially injected 5 K helium gas atoms will cool the diamond in less than 0.1 s according to kinetic theory calculations. These atoms move at ~160 m/s so ~$10^{-11}$ kg of helium will collide with the diamond in 0.1 s. The heat capacity of this helium is over $10^{-8}$ J/K while the heat capacity of a 1 µm diameter diamond is only $10^{-12}$ J/K. Diamond has a very high thermal conductivity so the centre and the outside of the diamond will remain at thermal equilibrium.

Before it reaches the inhomogeneous magnetic field, the NV- must be oriented to this inhomogeneous field so as to avoid torques, to ensure a predictable ODMR spectrum and to maximise the force experienced. A homogeneous magnetic field of 50 mT or more would provide a torque to align the NV- axis along this homogeneous magnetic field [26].

Dropping the diamond from a magnetogravitational trap could be achieved by applying an electric field to pull the diamond lower until it falls out of the bottom of the trap [25]. This would use the electric dipole induced in the diamond as the diamond must be electrically neutral before the superposition experiment to minimise unwanted electrical forces. A vertical hole would be drilled through the permanent magnet in reference [25] providing a homogeneous magnetic field. Dropping the diamond from an ion trap would be possible by neutralising the particle, but there would be a risk that the diamond would fall while it has just one or two electrical charges remaining. A very stiff trap would be required to trap a 1 µm diamond holding just one electrical charge.

Kinetic theory describes the effusion of gas through the small hole, following Graham's law [36]. The effusion flux (moles per second), $J = n v A/4$ for a molar density $n$ (for the gas at higher pressure going into a chamber at much lower pressure), a mean gas speed $v$, and an infinitesimally thin orifice of area $A$. The same equation describes both the flux out of the upper trap chamber (into the lower UHV chamber) and also out of the lower UHV chamber (into the pump). To maintain the pressure in the UHV chamber, the flux out must equal the flux in. The gas speed is the same for both effusions assuming the temperature remains at 5 K due to the entire experiment being held at this temperature inside a cryostat. Therefore $n_{trap} A_{trap} = n_{UHV} A_{UHV}$ which leads to $P_{trap} r^2_{trap} = P_{UHV} r^2_{UHV}$ for pressure in the higher pressure region $P$ and hole radius $r$ (using molar density = P/RT from the ideal gas law, and $A = \pi r^2$). A convenient radius for the holes is $r_{trap}$ = 25 µm and $r_{UHV}$ = 80 mm allowing $P_{trap}/P_{UHV} = 10^7$. The pressure in the trapping region has been taken to 7 x $10^{-8}$ mbar for a magnetogravitational diamond trap [25] which would permit pressures below $10^{-14}$ mbar in the UHV chamber.

The inhomogeneous magnetic field must be very stable run-to-run, so this should come from two opposing superconducting electromagnets in persistent mode, and two sharp pole pieces to concentrate the field. The pole pieces would be machined from Hyperco (an alloy of iron, cobalt and vanadium) and by applying enough magnetic field, above 2.4 T, these would be saturated avoiding Barkhausen noise. Superconducting NMR magnets in persistent mode are stable to better than one part in $10^9$ over a period of days. To estimate the magnetic inhomogeneity, we can take a sphere with radius 20 µm to represent the sharp edge of the pole piece (it could easily be sharper in practice). The magnetic field on the surface of this is 2.4 T coming from the superconducting magnets. 80 µm away from this, the magnetic inhomogeneity is over 5000 T/m, and with two such opposing magnets, the magnetic inhomogeneity, $dB/dx = 10^4$ T/m. Much larger values could be used by moving the magnets closer together, but this leaves a reasonable 160 µm gap to be maintained over a 1.5 m drop to aid machinability. Also this distance helps to reduce Casimir-Polder forces and electric patch potentials.

10,000 microwave π pulses are needed for the dynamic decoupling to reach spin coherence times above 0.5 s [29, 30]. Covering the full 1.5 m drop will require 150 antennas that are each 1 cm long. Each one should be 50 µm from the diamond to allow fast π pulses of ~50 ns with a reasonable microwave pulse power of ~10 W. Many high-power microwave switches would be used between the antennas to avoid requiring 150 microwave amplifiers.

The traditional room temperature spin readout of the NVC has a single-shot signal-to-noise-ratio (SNR) of only 0.03, so would require over $10^5$ nanodiamonds to be dropped just to get a single data point with SNR = 10 [37]. However, by keeping the diamond at 5 K, single-shot single-spin readout can be used via photoluminescence excitation (PLE) with excitation near to the zero phonon line at 637 nm. 95% NV- readout fidelity has been demonstrated [30, 38].

The superposition distance created is $s = ½ a t^2$ with $m a = g \mu_B dB/dx$ for a diamond of mass $m$, and an NV- centre with g-factor $g$ in a magnetic field with inhomogeneity $dB/dx$; $\mu_B$ is the Bohr magneton. After a drop time of 0.2 s we have over 2 μm superposition distance, leaving another 0.2 s for the two superposition components to be brought back together for matter-wave interference. The 0.4 s time is shorter than demonstrated NV- spin coherence times [29, 30].

Decoherence of the matter wave can come from gas atoms and from blackbody radiation. It has been shown that 5 K is cold enough that blackbody radiation will not cause wavefunction collapse even up to much larger superposition distance [39]. Very low pressures of around $5 \times 10^{-15}$ mbar would be needed to avoid decoherence from gas atoms as a single collision would be enough to collapse the superposition [39].

To observe interference fringes as evidence of the superposition, it has been suggested that the experiment could be tilted to create a gravitational phase [9, 11]. This phase is independent of the mass of the microdiamond which conveniently permits comparisons between drops with different microdiamonds. This gravitational phase is proportional to the drop time cubed, so varying this time would be another good way to observe fringes. Applying an electric field to one arm of the interferometer could be used to detect fringes but this phase would be dependent on the diamond mass so is not so convenient. All of these phase changes are very large for the parameters described here, which would lead to a very sensitive sensor if the phase were controlled.

In fact, a generic obstacle to any such large superpositions as are described in this proposal is that the phases developed are so large that they can be pseudo-random unless the environment is controlled precisely. The phase is an energy divided by $\hbar$, and $\hbar$ is so small that this phase is extremely sensitive to even very small energies. The phase was not problematically large for some of the smaller superpositions proposed previously [9].

To avoid this problem while having such large superpositions as proposed here, it is crucial for the inhomogeneous magnetic field ($dB/dx$) to be very horizontal with respect to gravity. This means that θ in references [9] and [11] must be very close to π/2 radians. If this is true to one part in $10^9$, the gravitational phase is $\sim 10^5$ rad for our parameters. Such control of θ would be achieved by measuring with laser interferometry the separation between the floor and a 1 m rigid arm sticking out the side of the experiment. This phase of $\sim 10^5$ rad will not be pseudo-random as long as the drop time, the magnitude of $dB/dx$, and the g-factor of the NV- are all controlled to one part in $10^5$. It is easy for the drop time to be controlled this well as it should be 0.4 s ±4 μs and we typically apply microwave pulses with ns precision. The relative magnitude of the magnetic field inhomogeneity between different drops will be controlled to better than 1 ppm using superconducting magnets in persistent mode and magnetically saturated pole pieces. To maintain the g-factor of the NV- centre drop-to-drop it will be important that its orientation does not change too much during the drop.

In conclusion, we have described the practicalities of putting a 1 μm-diameter diamond into a superposition with a spatial separation of over 1 μm, to test the limits of quantum mechanics.

[1]     E. Schrödinger, Naturwissenschaften **23**, 807 (1935).
[2]     C. Monroe, D. M. Meekhof, B. E. King and D. J. Wineland, Science **272**, 1131 (1996).
[3]     A. Bassi *et al.*, Reviews of Modern Physics **85**, 471 (2013).
[4]     A. Bassi and G. C. Ghirardi, Phys. Rep.-Rev. Sec. Phys. Lett. **379**, 257 (2003).

[5] S. Nimmrichter and K. Hornberger, Physical Review Letters **110**, 160403 (2013).
[6] P. Haslinger *et al.*, Nature Physics **9**, 144 (2013).
[7] S. Eibenberger *et al.*, Phys. Chem. Chem. Phys. **15**, 14696 (2013).
[8] Z.-q. Yin, T. Li, X. Zhang and L. M. Duan, Physical Review A **88**, 033614 (2013).
[9] M. Scala *et al.*, Physical Review Letters **111**, 180403 (2013).
[10] C. Wan *et al.*, Physical Review A **93**, 043852 (2016).
[11] C. Wan *et al.*, Physical Review Letters **117**, 143003 (2016).
[12] S. Bose *et al.*, Physical Review Letters **119**, 240401 (2017).
[13] R. J. Marshman *et al.*, arXiv:1807.10830 (2018).
[14] T. M. Hoang, J. Ahn, J. Bang and T. Li, Nature Communications **7**, 12250 (2016).
[15] L. P. Neukirch, E. von Haartman, J. M. Rosenholm and A. Nick Vamivakas, Nat. Photonics **9**, 653 (2015).
[16] R. M. Pettit, L. P. Neukirch, Y. Zhang and A. Nick Vamivakas, J. Opt. Soc. Am. B **34**, C31 (2017).
[17] L. P. Neukirch *et al.*, Opt. Lett. **38**, 2976 (2013).
[18] A. T. M. A. Rahman *et al.*, Scientific Reports **6**, 21633 (2016).
[19] A. C. Frangeskou *et al.*, New J. Phys. **20**, 043016 (2018).
[20] T. Delord, L. Nicolas, L. Schwab and G. Hetet, New J. Phys. **19**, 10 (2017).
[21] T. Delord, L. Nicolas, M. Bodini and G. Hétet, Appl. Phys. Lett. **111**, 013101 (2017).
[22] T. Delord *et al.*, Physical Review Letters **121**, 053602 (2018).
[23] A. Kuhlicke, A. W. Schell, J. Zoll and O. Benson, Appl. Phys. Lett. **105**, 0731012014).
[24] G. P. Conangla, A. W. Schell, R. A. Rica and R. Quidant, Nano Lett. **18**, 3956 (2018).
[25] J.-F. Hsu, P. Ji, C. W. Lewandowski and B. D'Urso, Scientific Reports **6**, 30125 (2016).
[26] Y. Ma *et al.*, Physical Review A **96**, 023827 (2017).
[27] T. M. Hoang *et al.*, Physical Review Letters **117**, 123604 (2016).
[28] A. M. Edmonds *et al.*, Physical Review B **86**, 035201 (2012).
[29] N. Bar-Gill *et al.*, Nature Communications **4**, 1743 (2013).
[30] M. H. Abobeih *et al.*, Nature Communications **9**, 2552 (2018).
[31] D. E. Chang *et al.*, Proceedings of the National Academy of Sciences **107**, 1005 (2010).
[32] A. T. M. A. Rahman, A. C. Frangeskou, P. F. Barker and G. W. Morley, Rev. Sci. Instrum. **89**, 023109 (2018).
[33] G. Ranjit, M. Cunningham, K. Casey and A. A. Geraci, Physical Review A **93**, 053801 (2016).
[34] D. C. Moore, A. D. Rider and G. Gratta, Physical Review Letters **113**, 251801 (2014).
[35] K. Fang *et al.*, Physical Review Letters **110**, 130802 (2013).
[36] E. A. Mason and B. Kronstadt, Journal of Chemical Education **44**, 740 (1967).
[37] D. Hopper, H. Shulevitz and L. Bassett, Micromachines **9**, 437 (2018).
[38] L. Robledo *et al.*, Nature **477**, 574 (2011).
[39] O. Romero-Isart, Physical Review A **84**, 052121 (2011).